\input harvmac
\input epsf

\overfullrule=0pt
\abovedisplayskip=12pt plus 3pt 
\belowdisplayskip=14pt plus 2pt 
%

\def\bar{\overline}
\def\to{\rightarrow}

\font\zfont = cmss10 
\font\litfont = cmr6

\def\bigone{\hbox{1\kern -.23em {\rm l}}}
\def\ZZ{\hbox{\zfont Z\kern-.4emZ}}
\def\half{{\litfont {1 \over 2}}}
\def\uone{\underline{1}}
\def\utwo{\underline{2}}
\def\uthree{\underline{3}}
\def\ufour{\underline{4}}
\def\ufive{\underline{5}}
\def\onefour{{\litfont{1\over 4}}}
\def\ua{{\underline{a}}}
\def\ub{{\underline{b}}}

\def\bw{{\bar w}}
\def\bu{{\bar u}}
\def\bv{{\bar v}}


\lref\bsv{R. Britto-Pacumio, A. Strominger and A. Volovich, {\it
``Holography for Coset Spaces''}, hep-th/9905211.}
\lref\hashit{A. Hashimoto and N. Itzhaki, {\it ``Non-commutative
Yang-Mills and the AdS/CFT Correspondence''}, hep-th/9907166.}
\lref\maldarus{J. Maldacena and J. Russo, {\it ``Large N Limit of
Non-commutative Gauge Theories''}, hep-th/9908134.}
\lref\topblack{R.B. Mann, {\it ``Topological Black Holes: Outside
Looking In''}, gr-qc/9709039\semi
M. Ba\~nados, A. Gomberoff and C. Martinez, {\it ``Anti-deSitter Space
and Black Holes''}, Class. Quant. Grav. {\bf 15} (1998) 3575;
hep-th/9805087.} 
\lref\hor{G. Horowitz and D. Marolf, {\it ``A New Approach to String
Cosmology''}, hep-th/9805207.}
\lref\gao{Y.-H. Gao, {\it ``AdS/CFT Correspondence and Quotient Space
Geometry''}, hep-th/9908134.} 
\lref\btz{M. Ba\~nados, C. Teitelboim and J. Zanelli, {\it ``The Black
Hole in Three-dimensional Space-time''}, Phys.Rev.Lett. {\bf 69} 
(1992) 1849; hep-th/9204099\semi
M. Ba\~nados, M. Henneaux, C. Teitelboim and J. Zanelli, {\it
``Geometry of the $2+1$ Black Hole''}, Phys. Rev. {\bf D48} (1993)
1506; gr-qc/9302012.}
\lref\blust{K. Behrndt and D. L\"ust, {\it ``Branes, Waves and AdS
Orbifolds''}, hep-th/9905180.}
\lref\kachsil{S. Kachru and E. Silverstein, {\it ``4-D Conformal
Theories and Strings on Orbifolds''}, Phys. Rev. Lett. {\bf 80} 
(1998) 4855; hep-th/9802183.}
\lref\maldacena{J. Maldacena, {\it ``The Large N Limit of
Superconformal Field Theories and Supergravity''},
Adv. Theor. Math. Phys. {\bf 2} (1998) 231; hep-th/9711200.}
\lref\acharya{B.S. Acharya, J. Figueroa O'Farrill, C. Hull and 
B. Spence, {\it ``Branes at Conical Singularities and Holography''}, 
Adv. Theor. Math. Phys. {\bf 2} (1999) 1249; hep-th/9808014.}
\lref\morples{D. Morrison and R. Plesser, {\it ``Non-Spherical 
Horizons-I''}, Adv. Theor. Math. Phys. {\bf 3} (1999) 1;
hep-th/9810201.}
\lref\lupopetown{H. L\"u, C.N. Pope and P.K. Townsend, 
{\it ``Domain Walls from Anti-deSitter Space-time''},
Phys. Lett. {\bf B391} (1997) 39; hep-th/9607164.}
\lref\klebwit{I. Klebanov and E. Witten, {\it ``Superconformal Field 
Theory on Threebranes at a Calabi- Yau Singularity''}, Nucl.Phys.
{\bf B536} (1998) 199; hep-th/9807080.}
\lref\uranga{A. Uranga, {\it ``Brane configurations for Branes at
Conifolds''}, JHEP {\bf 01} (1999) 022; hep-th/9811004.}
\lref\dasm{K. Dasgupta and S. Mukhi, {\it ``Brane Constructions,
Conifolds and M-Theory''}, Nucl. Phys. {\bf B551} (1999) 204; 
hep-th/9811139.}
\lref\tseytlin{A.A. Tseytlin, {\it ``Type IIB Instanton as a Wave in
Twelve Dimensions''}, Phys. Rev. Lett. {\bf 78} (1997) 1864;
hep-th/9612164.} 
\lref\popewar{C.N. Pope and N.P. Warner, {\it ``Two New Classes of
Compactifications of $d=11$ Supergravity''}, Class. Quant. Grav 
{\bf 2} (1985) L1.}

{\nopagenumbers
\Title{\vtop{\hbox{hep-th/9908192}
\hbox{TIFR/TH/99-44}}}
{\vtop{
\centerline{Killing Spinors and}
\medskip
\centerline{Supersymmetric AdS Orbifolds}}}
\centerline{Bahniman Ghosh\foot{E-mail: bghosh@theory.tifr.res.in}
and Sunil Mukhi\foot{E-mail: mukhi@tifr.res.in}}
\vskip 5pt
\centerline{\it Tata Institute of Fundamental Research,}
\centerline{\it Homi Bhabha Rd, Mumbai 400 005, India}
\vskip 5pt

\bigskip\bigskip

\centerline{ABSTRACT}
\medskip

We examine the behaviour of Killing spinors on $AdS_5$ under various
discrete symmetries of the spacetime. In this way we discover a number
of supersymmetric orbifolds, reproducing the known ones and adding a
few novel ones to the list. These orbifolds break the $SO(4,2)$
invariance of $AdS_5$ down to subgroups. We also make some comments on 
the non-compact Stiefel manifold $W_{4,2}$.

\Date{August 1999}
\vfill\eject}
\ftno=0

\newsec{Introduction}

Supersymmetric compactifications of type IIB string theory on
spacetimes of the form $AdS_5\times X_5$ have yielded a number of
interesting results about superconformal gauge theory in four
dimensions following the discovery\refs\maldacena\ of the AdS/CFT
correspondence. Here the compact manifold $X_5$ can be the sphere
$S^5$, or one of many possible ``non-spherical horizons'' including
spherical orbifolds and other Einstein spaces like the conifold base.

The presence of an $AdS_5$ factor guarantees conformal invariance of
the dual field theory, while varying the $X_5$ affects the spectrum of
the field theory and in particular the total number of
supersymmetries. Recently a number of situations have been discussed
where instead one keeps $X_5$ fixed (for example, to be $S^5$) and
chooses different noncompact Einstein spaces in lieu of $AdS_5$. Some
examples of this are the five-dimensional Stiefel manifold
$W_{4,2}$\refs\bsv, spaces with a nontrivial
$H=dB$\refs{\hashit,\maldarus}\ and orbifolds of
$AdS_p$\refs{\topblack,\hor,\gao,\blust}. The physical interpretation
of all these spaces is not completely clear at present -- for example,
some of the spaces discussed in \refs{\bsv,\hashit,\maldarus} have
singular behaviour at infinity leading to boundaries of dimension less
than 4.

Orbifolds of $S^5$\refs\kachsil\ are interesting because they allow us
to design spacetimes that are dual to a wide class of conformally
invariant, supersymmetric field theories in 4 dimensions. Some
orbifolds of $AdS_5$ have been interpreted as topological black
holes\refs\topblack\ generalizing the famous BTZ black hole in 3
dimensions\refs\btz, while other orbifolds represent cosmological
solutions\refs\hor\foot{Much of the previous work on $AdS$ orbifolds
deals with non-supersymmetric cases, and hence our discussion below
will not be closely related to it.}. A particular supersymmetric
$AdS_5$ orbifold was discussed in Ref.\refs\blust\ where it was
proposed to be dual to a 3-brane field theory with a $pp$-wave
propagating on it. Hence in this example one finds a type IIB
supergravity background that is dual to a 3-brane worldvolume theory,
not in its ground state but in a BPS excited state. This is an
intriguing direction in which to generalize the AdS/CFT
correspondence.

The purpose of this note is to examine conditions under which
orbifolds of $AdS_5$ (with or without fixed points) preserve some
supersymmetry. The analogous conditions for $S^5$ have been analyzed
in some depth in Refs.\refs{\acharya,\morples}. One key result that
helped in that classification was a theorem relating Killing spinors
on Einstein 5-manifolds to parallel spinors on a 6d cone above
them. However, one could also reproduce many of those results by
directly studying the transformation properties of Killing spinors
on $S^5$ under the orbifolding action. 

In what follows, we construct Killing spinors on $AdS_5$ in three
different coordinate systems and examine their behaviour under various
possible orbifolding actions. This enables us to construct a number of
supersymmetric orbifolds, including some known ones and some that are
apparently new. As we will see, different orbifolds can be
conveniently studied in different coordinates systems on $AdS_5$. A
complete classification of orbifolds of $AdS_5$, on the lines of
Refs.\refs{\acharya,\morples}, would be interesting to attempt. This
would perhaps follow if one could prove a theorem relating Killing
spinors on a (non-compact) 5-dimensional Einstein space to spinors on
a ``cone'' over it with two timelike directions.

We also compute Killing spinors for $W_{4,2}$ and discuss some
supersymmetric orbifolds of this space. The physical meaning of this
space and its relevance to the AdS/CFT correspondence are not very
clear, and the same holds for the $AdS$ orbifolds we consider except
in a few cases. We leave the detailed analysis of this question, along
with the study of the global structure of these orbifold spacetimes,
for the future.

Like the cases discussed in Refs.\refs{\blust,\bsv}, the orbifolds
discussed here break the $SO(4,2)$ invariance of $AdS_5$ down to
subgroups, while preserving the $S^5$ factor and hence the $SO(6)$
symmetry associated to R-symmetry of the boundary CFT. One can of
course combine the orbifolds discussed here with the ones proposed in 
Ref.\refs{\kachsil}\ to get compactifications with still lower
symmetry and supersymmetry.

\newsec{Killing Spinors on $AdS^5$}

$AdS^5$ spacetime can be described as a hyperboloid in a 6-dimensional 
spacetime with 2 timelike directions. Labelling the coordinates of
this ambient spacetime as $X_{-1}, X_0, X_1,\ldots X_4$, the metric is
\eqn\ambmet{
ds^2 = - (dX_{-1})^2 - (dX_0)^2 + (dX_1)^2 + \ldots + (dX_4)^2 
}
and the equation of the hyperboloid is:
\eqn\hyp{
-1 = - (X_{-1})^2 - (X_{0})^2 + (X_{1})^2 + (X_{2})^2 + (X_{3})^2 +
(X_{4})^2
}
The metric on $AdS^5$ is the one induced from the ambient space.
 
We will find it convenient to work in three different sets of
coordinates. 

\subsec{Light-Cone Type Coordinates}

These consist of two pairs of lightlike coordinates and one complex
coordinate. It is defined by
\eqn\lightcoor{
z_{1}^{\pm}=X_{0} \pm X_{1},\quad z_{2}^{\pm}=X_{2} \pm X_{-1},\quad w
= (X_3 + i X_4) } 
and the hyperboloid is
\eqn\lighthyp{
-1 = - z_1^+ z_1^- + z_2^+ z_2^- + w \bw
}
For this set of coordinates, it is convenient to choose an explicit
basis for the Gamma-matrices as:
\eqn\lightgamma{
\Gamma_1 = \sigma_2 \otimes 1,\quad 
\Gamma_2 =i \sigma_3 \otimes \sigma_1,\quad
\Gamma_3 = \sigma_3 \otimes \sigma_2,\quad
\Gamma_4 = \sigma_3 \otimes \sigma_3,\quad
\Gamma_5 = \sigma_1 \otimes 1\quad}
Note that $(\Gamma_2)^2 = -1$, while the other $\Gamma$-matrices
square to $+1$.

From the light-cone type coordinates we go to a set of five
independent coordinates $\theta_1,\theta_2,\alpha,\beta,\delta$ where
$0\le \theta_2 \le \pi$, $0\le \beta\le 2\pi$ and $\alpha$,$\delta$
and $\theta_1$ are non-compact. These coordinates are defined by:
\eqn\lightangles{
\eqalign{
z_1^{\pm} &= \cosh{\theta_1\over 2}e^{\pm\delta}\cr
z_2^{\pm} &= \sinh{\theta_1\over 2}\, \cos{\theta_2\over
2}e^{\pm\alpha}\cr 
w &= \sinh{\theta_1\over 2}\, \sin{\theta_2\over
2}e^{i\beta}\cr}}
By abuse of notation we will refer to these also as light-cone type
coordinates, though they are actually an angular parametrization of
those coordinates which solves the hyperboloid constraint. The metric
on $AdS^5$ in these coordinates is:
\eqn\lightmet{
ds^2 = -\sinh^2{\theta_1\over 2} \cos^2{\theta_2\over 2} d\alpha^2 +
\cosh^2{\theta_1\over 2} d\delta^2 +  \onefour d\theta_1^2 + 
\sinh^2{\theta_1\over 2}(\onefour
d\theta_2^2 + \sin^2{\theta_2\over 2} d\beta^2 )}
For fixed $\theta_2$ and $\beta$, the metric is proportional to that
of $AdS^3$.

For this case, we have the vielbeins:
\eqn\lightviel{
\eqalign{
&e^{\uone} = \half\,\sinh{\theta_1\over2} d\theta_2\qquad
e^{\utwo} = \sinh{\theta_1\over2} \, \cos{\theta_2\over 2}\, d\alpha\cr
&e^{\uthree} = \cosh{\theta_1\over2}\, d\delta\qquad
e^{\ufour} = \half\, d{\theta_1}\qquad
e^{\ufive} = \sinh{\theta_1\over2}\, \sin{\theta_2\over2}\, d\beta\cr }}
and the spin connections:
\eqn\lightspin{
\eqalign{
&\omega^{\uone\utwo} = \sin{\theta_2\over 2}\, d\alpha\qquad
\omega^{\uone\ufour} = \half \cosh{\theta_1\over 2}\, d\theta_2\qquad
\omega^{\uone\ufive} = - \cos{\theta_2\over 2} d\beta\cr
&\omega^{\utwo\ufour} = \cos{\theta_2\over 2}\, \cosh{\theta_1\over
2}\, d\alpha\qquad
\omega^{\uthree\ufour} = \sinh{\theta_1\over 2}\, d\delta\qquad
\omega^{\ufour\ufive} = - \sin{\theta_2\over 2}\, \cosh{\theta_1\over
2}\, d\beta\cr}}
Note that the tangent-space metric has $\eta_{\utwo\utwo}=-1$ while
the other components are $+1$. 

We are interested in studying the Killing spinors on $AdS^5$ in this
coordinate basis. The relevant equation is:
\eqn\killspin{
(\del_\mu + {1\over 4}\omega_\mu^{\ua\ub}\Gamma_{\ua\ub} - {1\over
2} e_\mu^\ua \Gamma_\ua) \epsilon =0
}
It is fairly straightforward to compute the solutions to this 
equation, which are given by:
\eqn\lightkill{
\epsilon = e^{{1\over4}\Gamma_4\theta_1} 
e^{-{1\over4}\Gamma_{14}\theta_2}
e^{-{1\over2}\Gamma_{24}\alpha}
e^{{1\over2}\Gamma_3\delta}
e^{\half\Gamma_{15}\beta} \epsilon_0 } 
where $\epsilon_0$ is an arbitrary constant spinor. 

\subsec{Complex Coordinates}

Another coordinate system will turn out to be useful to investigate a
different class of orbifolds. These will be called complex coordinates
-- they are actually complex coordinates of the ambient 6-dimensional
spacetime, a complex time and two complex space dimensions. Thus we
define:
\eqn\compcoor{
u=X_{-1} + iX_0,\quad v =X_{1} + i X_2,\quad w = (X_3 + i X_4) }
in terms of which the hyperboloid is
\eqn\comphyp{
-1 = -u\bu + v\bv + w\bw }
The coordinate $w$ is the same as was used for the light-cone type
coordinates. This time it is convenient to go to five independent
coordinates $\theta_1,\theta_2,\alpha',\beta,\delta'$ where $0\le
\theta_2 \le \pi$, $0\le \beta,\alpha',\delta'\le 2\pi$ and $\theta_1$
is non-compact. These coordinates are defined by:
\eqn\compangles{
\eqalign{
u &= \cosh{\theta_1\over 2}e^{i\delta'}\cr
v &= \sinh{\theta_1\over 2}\, \cos{\theta_2\over
2}e^{i\alpha'}\cr 
w &= \sinh{\theta_1\over 2}\, \sin{\theta_2\over
2}e^{i\beta}\cr}}
Again, although these angles parametrize the complex coordinates in a
way which solves the hyperboloid constraint, we will refer to the
angles themselves as complex coordinates. 

Note that these coordinates can be obtained from the previous ones by
the formal replacement $\alpha=i\alpha'$ and $\delta=i\delta'$, which
interchanges one space with one time direction. This replacement in
Eq.\lightmet\ also gives us the metric in these coordinates. It is
evident that the vielbeins are formally the {\it same} as before,
though the tangent-space metric now has
$\eta_{\uthree\uthree}=-1$. Careful inspection shows that the spin
connections also turn out to be exactly the same as in Eq.\lightspin,
as the sign changes introduced by the interchange of a space and a
time direction eventually cancel out. The change of tangent space
metric necessitates a slightly different basis of
$\Gamma$-matrices. We multiply $\Gamma_2$ of Eq.\lightgamma\ by $-i$
and $\Gamma_3$ by $i$. Hence the new set of $\Gamma$-matrices (which
we label $\Gamma'$ to avoid confusion with the previous set) becomes:
\eqn\compgamma{
\Gamma'_1 = \sigma_2 \otimes 1,\quad 
\Gamma'_2 = \sigma_3 \otimes \sigma_1,\quad
\Gamma'_3 =i \sigma_3 \otimes \sigma_2,\quad
\Gamma'_4 = \sigma_3 \otimes \sigma_3,\quad
\Gamma'_5 = \sigma_1 \otimes 1\quad}
This time, $(\Gamma'_3)^2=-1$ while the others square to $+1$. The
advantage of this choice is that we find (formally) the same Killing
spinor as in Eq.\lightkill, but now with the $\Gamma'$-matrices:
\eqn\compkill{
\epsilon = e^{{1\over4}\Gamma'_4\theta_1} 
e^{-{1\over4}\Gamma'_{14}\theta_2}
e^{-{1\over2}\Gamma'_{24}\alpha}
e^{{1\over2}\Gamma'_3\delta}
e^{\half\Gamma'_{15}\beta} \epsilon_0 }

\subsec{Horospherical Coordinates}

Let us finally recall\refs\lupopetown\ the Killing spinors in
horospherical coordinates, which consist of five independent real
coordinates, $r,x^1,x^2,x^3,x^4$ in terms of which the metric is:
\eqn\killmet{
ds^2 = (dr)^2 + e^{2r}(-(dx^1)^2 + (dx^2)^2 +(dx^3)^2 + (dx^4)^2 )
}
Choosing the $ \Gamma $ matrices as
\eqn\ga{
\Gamma_1 = i\sigma_3 \otimes \sigma_2,\quad
\Gamma_2 = \sigma_3 \otimes \sigma_1,\quad
\Gamma_r = \sigma_3 \otimes \sigma_3,\quad
\Gamma_3 = \sigma_1 \otimes 1,\quad
\Gamma_4 = \sigma_2 \otimes 1\quad
}
the Killing spinor is found to be\refs\lupopetown:
\eqn\horokill{
\eqalign{
\epsilon &= e^{{1\over 2} r \Gamma_{r}}(1 + {1\over 2} x^{\alpha}
\Gamma_{\alpha}(1-\Gamma_{r}))\epsilon_0\cr
&= \pmatrix{ e^{r \over 2}(\epsilon_0^{(1)} + x^{+}\epsilon_0^{(2)} 
+ (x^3 - i x_4)\epsilon_0^{(3)})\cr
\phantom{a}\cr
e^{-{r \over 2}}\epsilon_0^{(2)}\cr
\phantom{a}\cr
e^{-{r \over 2}}\epsilon_0^{(3)}\cr
\phantom{a}\cr
e^{r \over 2}(\epsilon_0^{(4)} + x^{-}\epsilon_0^{(3)} + (x^3 + i
x^4)\epsilon_0^{(2)})\cr}}
}
where $\alpha=1,2,3,4$, $x^{+}=x^{1} + x^{2}$, 
$x^{-}=x^{1} - x^{2}$ and
\eqn\epsconst{
\epsilon_{0} = \pmatrix{\epsilon_0^{(1)}\cr
\epsilon _0^{(2)}\cr \epsilon_0^{(3)}\cr \epsilon_0^{(4)}\cr}
}
The transformation between these horospherical coordinates and the
light-cone type coordinates $z_1^\pm, z_2^\pm, w$ is:
is
\eqn\horltrans{
e^r = z_{2}^{+},\quad
x^{+} = {z_{1}^{+}\over z_{2}^{+}},\quad
x^{-} = {z_{1}^{-}\over z_{2}^{+}},\quad
x^{3} = {(w + \bar w)\over 2 z_{2}^{+}},\quad
x^{4} = {(w - \bar w)\over 2 i z_{2}^{+}},\quad
}

\newsec{Orbifolds of $AdS_5$}

\subsec{Half-supersymmetric Orbifolds}

Now we will examine the orbifolding actions which are natural in the
various coordinates. In the light-cone type coordinates, one natural
action follows from the following transformation:\refs\blust:
\eqn\lightorb{
z_1^{\pm} \to e^{\pm 2\pi /k} z_1^{\pm},\qquad 
z_2^{\pm} \to e^{\pm 2\pi /k} z_2^{\pm} }
which can be expressed as a simple translation:
\eqn\lighttrans{
\delta \to \delta + {2\pi\over k},\qquad
\alpha \to \alpha + {2\pi\over k}
 }
From Eq.\lighthyp, this clearly has no fixed points, since the
hyperboloid does not pass through $z_1^\pm=z_2^\pm=0$. 
In order for the Killing spinor $\epsilon$ to be invariant under the
above transformation we must require that the constant spinor
$\epsilon_0$ satisfy
\eqn\epscond{
e^{{\pi\over k}(-\Gamma_{24} + \Gamma_3)} \epsilon_0 = \epsilon_0}
which means the matrix $(-\Gamma_{24} + \Gamma_3)$ must
annihilate $\epsilon_0$. In the basis chosen in Eq.\lightgamma, we
have:
\eqn\explmat{
-\Gamma_{24} + \Gamma_3 = 2 \pmatrix{0 &0\cr 0 & \sigma_2\cr} }
and hence the Killing spinors that are preserved, in this basis, are
the ones for which
\eqn\preserv{
\epsilon_0 = \pmatrix{\epsilon_0^{(1)}\cr \epsilon_0^{(2)}\cr 0\cr 0\cr}}
This in particular gives a direct proof that the orbifold discussed in 
Ref.\refs\blust\  preserves half the supersymmetries.

The above orbifold action is generated by an $SU(1,1)$ matrix in the
full isometry group $SO(4,2)$ of $AdS_5$, hence the surviving symmetry
group is the commutant of $SU(1,1)$ in $SO(4,2)$ which is
$SU(1,1)\times U(1)$. This is analogous to the fact that the simplest
half-supersymmetric orbifold of $S^5$ (corresponding to D3-branes at
an ALE singularity) has an R-symmetry group $SU(2)\times U(1)$. 

Turning now to the complex coordinates, it is natural to consider
orbifold actions of the type
\eqn\comporb{
u\to \gamma^d u,\qquad v\to \gamma^a v,\qquad w\to \gamma^b w
}
where $\gamma = \exp(2\pi i/k)$ and $a,b,d$ are some integers. These
are quite analogous to corresponding orbifolds of $S^5$. The result is 
also analogous: the orbifolding action above leaves the Killing spinor 
invariant if
\eqn\compcond{
\left(-a \Gamma'_{24} + d \Gamma'_3 + b \Gamma'_{15}\right)\epsilon_0
= 0
}
The above matrix has eigenvalues $(a+b-d)$, $-(a+b+d)$, $(a-b+d)$ and
$-(a-b-d)$. If one of $a,b,d$ is zero then we have two vanishing
eigenvalues and $\half$-supersymmetry.

Note, however, that because of the signature of the spacetime, all the
$\half$-supersymmetric cases are not equivalent. The case with $d=0$ has
a circle of fixed points $u\bu=1$, while the cases with $a=0$ or $b=0$
have no fixed points and are equivalent to each other.

In the case $d=0$, the orbifold generator lies in an $SU(2)$ subgroup
of $SO(4)\subset SO(4,2)$ hence the symmetry of the quotient space is
$SU(2) \times U(1)$. On the other hand, for $a=0$ or $b=0$ the
orbifold is generated by an element in an $SU(1,1)$ subgroup of $SO(2,2)
\subset SO(4,2)$ and the surviving symmetry is $SU(1,1)\times
U(1)$. 

Next, it is useful to examine the orbifold described in Eq.\lightorb,
in horospherical coordinates. The action becomes:
\eqn\hororb{
r \to r + a ,\quad
x^{+} \to x^{+} ,\quad
x^{-} \to e^{-2a} x^{-} ,\quad
x^{3} \to e^{-a} x^{3} ,\quad
x^{4} \to e^{-a} x^{4} ,\quad
}
Hence the Killing spinor transforms as
\eqn\hortrkill{
\epsilon \to e^{-{1\over 2} a (1 \otimes \sigma_3)}
\pmatrix{ e^{(r + a)\over
2}(\epsilon_0^{(1)} + x^{+}\epsilon_0^{(2)} + e^{-a}(x^3 - ix^4)
\epsilon_0^{(3)})\cr
\phantom{a}\cr
e^{-{(r+a)\over 2}} \epsilon_0^{(2)}\cr
\phantom{a}\cr
e^{-{(r+a)\over 2}} \epsilon_0^{(3)}\cr
\phantom{a}\cr
e^{(r + a)\over 2}(\epsilon_0^{(4)} + 
e^{-2a}x^{-}\epsilon_0^{(3)} + e^{-a}(x^3 +
ix^4)\epsilon_0^{(2)})\cr} 
}
Thus the Killing spinors that are preserved by this orbifold, in this
basis, are the ones for which
\eqn\horpres{
\epsilon_0 = \pmatrix{\epsilon_0^{(1)}\cr \epsilon_0^{(2)}\cr 0\cr
0\cr}}

Another apparently trivial kind of $\half$-supersymmetric orbifold is
apparent from Eqn.\horokill. Suppose we choose
$\epsilon_0^{(2)}=\epsilon_0^{(3)}=0$. Then the Killing spinor becomes
independent of $x^\pm,x^3,x^4$. As a result, periodic identifications
in these coordinates preserve the Killing spinor. This is essentially
what was noted in Ref.\lupopetown, and corresponds to the fact that
the identification of these coordinates breaks conformal invariance by
introducing a scale, hence the conformal part of the superconformal
invariance goes away. Thus such orbifolds preserve half the
supersymmetries. (One can further deform the space in the $x^3,x^4$
directions and add a nontrivial $B$-field, preserving the remaining
supersymmetry, as was done in Ref.\refs{\hashit,\maldarus}. In this
case one does not expect the deformed manifold to have a Killing
spinor, since the field strength $dB$ also contributes to the
supersymmetry variation.)

\subsec{One-fourth Supersymmetry}

We have already considered orbifold actions, in complex coordinates,
of the general type
\eqn\comporbtwo{
u\to \gamma^d u,\qquad v\to \gamma^a v,\qquad w\to \gamma^b w
}
where $\gamma = \exp(2\pi i/k)$ and $a,b,d$ are some integers. We saw
that the relevant matrix acting on Killing spinors has eigenvalues
$(a+b-d)$, $-(a+b+d)$, $(a-b+d)$ and $-(a-b-d)$. Hence if all of
$a,b,d$ are nonzero then at most one of these eigenvalues can be zero
and in that case we have a $1\over 4$-supersymmetric orbifold. For the
$1\over 4$ supersymmetric cases we have no fixed points for the
orbifold action.

The orbifold generator lies in an $SU(2,1)$ subgroup of $SO(4,2)$,
hence it preserves only a $U(1)$ symmetry. This is the analogue of the 
$U(1)$ symmetry preserved by ${1\over 4}$-supersymmetric orbifolds of
$S^5$, which is realized as a $U(1)$ R-symmetry in the boundary
theory. 

Another class of ${1\over 4}$-supersymmetric $AdS_5$ orbifolds comes
from quotienting by a pair of transformations each of which preserves
half the supersymmetry. If $(k,k')$ are co-prime there are two
inequivalent cases:
\eqn\firstcase{
\eqalign{
u \to \gamma u,\qquad &v\to \gamma^{-1} v\cr
v \to \gamma' v,\qquad & w\to \gamma'^{-1} w\cr}}
and
\eqn\secondcase{
\eqalign{
u \to \gamma u,\qquad &v\to \gamma^{-1} v\cr
u \to \gamma' u,\qquad & w\to \gamma'^{-1} w\cr}}
where $\gamma^k= (\gamma')^{k'} =1$.

Another interesting ${1\over 4}$-supersymmetric orbifold arises by
combining the periodic identification in $x^2,x^3,x^4$ in the
horospherical coordinates, with the orbifolding action of
Eq.\hororb. Here the constant Killing spinor satisfies
$\epsilon_0^{(2)} = \epsilon_0^{(3)}= \epsilon_0^{(4)}=0$.

We have encountered a number of supersymmetric orbifolds, but it turns
out that each of them is natural in a certain coordinate system and
not so easy to describe in another. Thus, it becomes hard to combine
the different actions discussed in the previous section and find more
general orbifolds. For example, one of the simplest orbifolds
described in complex coordinates in Eq.\comporb\ arises by choosing
$k=2, d=a=1, b=0$. This just corresponds to the reflection $u\to -u$,
$v\to -v$. In terms of the light cone type coordinates this means
$z_i^\pm \to - z_i^\pm$, which cannot be carried out using the
independent coordinates defined in Eq.\lightangles, which only cover
the region $z_1^\pm >0$.

In contrast, the orbifold corresponding to $k=2, a=b=1, d=0$ can be
realized in the light cone type coordinates. In this case we have
$v\to -v$, $w\to -w$. This corresponds to the action
\eqn\lightsimp{
z_1^+ \to z_1^-,\qquad z_2^+ \to - z_2^-,\qquad w\to -w
}
which in terms of the independent coordinates in Eq.\lightangles\ is
just 
\eqn\lightangsimp{
\delta \to -\delta,\qquad \alpha \to -\alpha,\qquad \theta_1 \to 
-\theta_1} 
This acts on the Killing spinor in Eq.\lightkill\ as follows.
In our basis, this Killing spinor is explicitly given by
\eqn\explkill{
\epsilon = \pmatrix{
\matrix{
\cos{\theta_2\over 4} e^{\theta_1\over 4}\left\{
\cosh{\alpha -\delta\over 2}\epsilon_0^{(1)} +
i\sinh{\alpha-\delta\over2}\epsilon_0^{(2)} \right\}e^{-i{\beta\over
2}}\cr
-i\sin{\theta_2\over 4} e^{\theta_1\over 4}\left\{
\cosh{\alpha +\delta\over 2}\epsilon_0^{(3)} +
i\sinh{\alpha+\delta\over2}\epsilon_0^{(4)} \right\}e^{i{\beta\over
2}}}\cr
\phantom{{a\over b}}\cr
\matrix{
\cos{\theta_2\over 4} e^{-{\theta_1\over 4}}\left\{
\cosh{\alpha -\delta\over 2}\epsilon_0^{(2)} -
i\sinh{\alpha-\delta\over2}\epsilon_0^{(1)} \right\}e^{-i{\beta\over
2}}\cr
+i\sin{\theta_2\over 4} e^{-{\theta_1\over 4}}\left\{
\cosh{\alpha +\delta\over 2}\epsilon_0^{(4)} -
i\sinh{\alpha+\delta\over2}\epsilon_0^{(3)} \right\}e^{i{\beta\over
2}}}\cr
\phantom{{a\over b}}\cr
\matrix{
\cos{\theta_2\over 4} e^{-{\theta_1\over 4}}\left\{
\cosh{\alpha +\delta\over 2}\epsilon_0^{(3)} +
i\sinh{\alpha+\delta\over2}\epsilon_0^{(4)} \right\}e^{i{\beta\over
2}}\cr
-i\sin{\theta_2\over 4} e^{-{\theta_1\over 4}}\left\{
\cosh{\alpha -\delta\over 2}\epsilon_0^{(1)} +
i\sinh{\alpha-\delta\over2}\epsilon_0^{(2)} \right\}e^{-i{\beta\over
2}}}\cr
\phantom{{a\over b}}\cr
\matrix{
\cos{\theta_2\over 4} e^{\theta_1\over 4}\left\{
\cosh{\alpha +\delta\over 2}\epsilon_0^{(4)} -
i\sinh{\alpha+\delta\over2}\epsilon_0^{(3)} \right\}e^{i{\beta\over
2}}\cr
+i\sin{\theta_2\over 4} e^{\theta_1\over 4}\left\{
\cosh{\alpha -\delta\over 2}\epsilon_0^{(2)} -
i\sinh{\alpha-\delta\over2}\epsilon_0^{(1)} \right\}e^{-i{\beta\over
2}}}\cr
}}
Now one finds that, setting
$\epsilon_0^{(1)}=\epsilon_0^{(2)}$ and $\epsilon_0^{(3)} =
-\epsilon_0^{(4)}$, the expression for $\epsilon$ above goes over to
$\epsilon'$ satisfying
\eqn\lightepstrans{
\epsilon' = \pmatrix{\sigma_1 &0\cr 0 & -\sigma_1\cr}\epsilon =
(\sigma_3 \otimes \sigma_1)\epsilon
}
This coincides with the Lorentz transformation of $\epsilon$ under the
action in Eq.\lightangsimp\foot{The action in Eq.\lightangsimp\ inverts the
sign of the vielbeins $e^{\uone}, e^{\uthree}, e^{\ufour}, e^{\ufive}$,
hence it must be represented on spinors by a matrix $P$ which
anticommutes with $\Gamma_{\uone},\Gamma_{\uthree},\Gamma_{\ufour},
\Gamma_{\ufive}$. Thus $P$ is proportional to $\Gamma_2=
i\sigma_3\otimes \sigma_1$. Since $P^2=1$, it follows that
$P=\sigma_3\otimes \sigma_1$.}

Thus we see (as expected from the analysis of the same orbifold in
complex coordinates) that this orbifold preserves half the
supersymmetries. But now that we know the preserved Killing spinors in
the basis appropriate to light-cone coordinates, we can combine this
orbifold with the orbifolding along lightlike directions defined in
Eq.\lightorb\ and determine if there is any residual
supersymmetry. Indeed we have seen that Eq.\lightorb\ preserves
Killing spinors of the form Eq.\lightkill\ where the constant spinor
$\epsilon_0$ has the form:
\eqn\onespin{
\epsilon_0= \pmatrix{\epsilon_0^{(1)}\cr \epsilon_0^{(2)} \cr 0\cr 0\cr}
}
while Eq.\lightangsimp\ preserves Killing spinors where $\epsilon_0$
has the form
\eqn\twospin{
\epsilon_0= \pmatrix{\epsilon_0^{(1)}\cr \epsilon_0^{(1)}\cr 
\epsilon_0^{(3)}\cr -\epsilon_0^{(3)} \cr}
}
Choosing now $\epsilon_0^{(1)}=\epsilon_0^{(2)}$ in the first of these
equations and $\epsilon_0^{(3)}=0$ in the second, we find that both
the transformations together preserve ${1\over 4}$ of the
supersymmetry, namely the Killing spinor for which
\eqn\commspin{
\epsilon_0 = \pmatrix{1\cr 1\cr 0\cr 0\cr}}
This is a $Z\times Z_2$ orbifold of $AdS_5$. It is much more
difficult, if not impossible, to express the $Z_k$ orbifold given in
complex coordinates by $v\to\gamma v, w\to \gamma^{-1}w, \gamma^k=1$
in terms of light-cone coordinates. Luckily it is not necessary to do
this. These orbifolds preserve the same Killing spinors for all
$k$. Since we have shown that for $k=2$ there is a ${1\over
4}$-supersymmetric $Z\times Z_2$ orbifold obtained by combining with
the action Eq.\lightorb, it follows that there is also a $Z\times Z_k$
orbifold with ${1\over 4}$-supersymmetry for all $k$.

\newsec{The Stiefel Manifold $W_{4,2}$}

Although not directly related to orbifolds of $AdS_5$, the noncompact
Stiefel manifold $W_{4,2}$\refs\bsv\ is an interesting ${1\over
4}$-supersymmetric coset spacetime. We will speculate later on its
possible relation with $AdS_5$ orbifolds. In the present section we
will compute its single Killing spinor and comment on orbifolds of
this spacetime.

This case corresponds to an analytic continuation of the compact
manifold variously denoted $T_{1,1}$ or $V_{4,2}$, which is the base
of the conifold geometry and hence appears in the study of D3-branes
at conifolds\refs\klebwit. It can be thought of as the coset space
$(AdS_3 \times AdS_3)/U(1)$, and also as a (timelike) $U(1)$ fibration
above {\it Euclidean} $AdS_2\times AdS_2$. The natural coordinates for
$W_{4,2}$ are $(\theta_1,\phi_1,\theta_2,\phi_2,\psi)$ where $0 \le
\psi < 4\pi$, $0 \le \phi_i < 2\pi$ while $\theta_i$ are
noncompact. In terms of these, the metric of $W_{4,2}$ is
\eqn\tonemet{
\eqalign{
ds^2 = &-{1\over 9}(d\psi + \cosh\theta_1\, d\phi_1 + \cosh\theta_2\,
d\phi_2)^2\cr
&+ {1\over 6}(d\theta_1^2 + \sinh^2 \theta_1\, d\phi_1^2)
+ {1\over 6}(d\theta_2^2 + \sinh^2 \theta_2\, d\phi_2^2)}}
which explicitly exhibits the $U(1)$ fibration, with $\psi$ being the
fibre coordinate. 

This spacetime breaks the maximal $SO(4,2)$ isometry to $SO(2,2)\times
SO(2)$, much as its compact version breaks the maximal $SO(6)$
isometry of $S^5$ to $SO(4)\times SO(2)$. From the metric one can
read off the vielbeins:
\eqn\conifviel{
\eqalign{
e^{\uone} &= {1\over\sqrt{6}} d\theta_1\qquad
e^{\utwo} = {1\over\sqrt{6}} \sinh\theta_1\, d\phi_1\cr
e^{\uthree} &= {1\over\sqrt{6}} d\theta_2\qquad
e^{\ufour} = {1\over\sqrt{6}} \sinh\theta_2\, d\phi_2\cr
e^{\ufive} &= {1\over 3}(d\psi + \cosh\theta_1\, d\phi_1 +
\cosh\theta_2\, d\phi_2)}}
and compute the spin connections:
\eqn\conifspin{
\eqalign{
\omega^{\uone\utwo} &= -{2\over 3}\cosh \theta_1\, d\phi_1
+ {1\over 3} d\psi + {1\over 3} \cosh \theta_2\, d\phi_2\cr
\omega^{\uone\uthree} &= \omega^{\uone\ufour} = \omega^{\utwo\uthree}
= \omega^{\utwo\ufour} = 0 \cr
\omega^{\uthree\ufour} &= -{2\over 3}\cosh \theta_2\, d\phi_2
+ {1\over 3} d\psi + {1\over 3} \cosh \theta_1\, d\phi_1\cr
\omega^{\uone\ufive} &= - {1\over\sqrt 6} \sinh\theta_1\, d\phi_1\qquad
\omega^{\utwo\ufive} =  {1\over\sqrt 6} d\theta_1\cr
\omega^{\uthree\ufive} &= - {1\over\sqrt 6} \sinh\theta_2\, d\phi_2\qquad
\omega^{\ufour\ufive} =  {1\over\sqrt 6} d\theta_2\cr}}
A convenient basis for the $\Gamma$-matrices this time is
\eqn\gamstief{
\Gamma_1 = \sigma_1 \otimes 1,\quad 
\Gamma_2 =\sigma_2 \otimes 1,\quad
\Gamma_3 = \sigma_3 \otimes \sigma_1,\quad
\Gamma_4 = \sigma_3 \otimes \sigma_2,\quad
\Gamma_5 = i\sigma_3 \otimes \sigma_3\quad}

In this basis, the Killing spinor equations become
\eqn\conifkillone{
{\del\over \del\theta_1}\pmatrix{\epsilon^{(1)}\cr \epsilon^{(2)}\cr 
\epsilon^{(3)}\cr \epsilon^{(4)}\cr} =
\pmatrix{0 & 0 & {1\over\sqrt 6} & 0\cr
0 & 0 & 0 & 0\cr
{1\over\sqrt 6} & 0 & 0 & 0\cr
0 & 0 & 0 & 0\cr}\pmatrix{\epsilon^{(1)}\cr \epsilon^{(2)}\cr 
\epsilon^{(3)}\cr \epsilon^{(4)}\cr} }
\eqn\conifkilltwo{
{\del\over \del\phi_1}\pmatrix{\epsilon^{(1)}\cr \epsilon^{(2)}\cr 
\epsilon^{(3)}\cr \epsilon^{(4)}\cr} =
\pmatrix{{i\over 3}\cosh\theta_1 & 0 & {i\over\sqrt 6}\sinh\theta_1 & 0\cr
0 & {i\over 3}\cosh\theta_1 & 0 & 0\cr
-{i\over\sqrt 6}\sinh\theta_1 & 0 & -{2i\over 3}\cosh\theta_1 & 0\cr
0 & 0 & 0 & 0\cr}\pmatrix{\epsilon^{(1)}\cr \epsilon^{(2)}\cr 
\epsilon^{(3)}\cr \epsilon^{(4)}\cr}}
\eqn\conifkillthree{
{\del\over \del\theta_2}\pmatrix{\epsilon^{(1)}\cr \epsilon^{(2)}\cr 
\epsilon^{(3)}\cr \epsilon^{(4)}\cr} =
\pmatrix{0 & {1\over\sqrt 6} & 0 & 0\cr
{1\over\sqrt 6} & 0 & 0 & 0\cr
0 & 0 & 0 & 0\cr
0 & 0 & 0 & 0\cr}\pmatrix{\epsilon^{(1)}\cr \epsilon^{(2)}\cr 
\epsilon^{(3)}\cr \epsilon^{(4)}\cr}}
\eqn\conifkillfour{
{\del\over \del\phi_2}\pmatrix{\epsilon^{(1)}\cr \epsilon^{(2)}\cr 
\epsilon^{(3)}\cr \epsilon^{(4)}\cr} =
\pmatrix{{i\over 3}\cosh\theta_2 & {i\over\sqrt 6}\sinh\theta_2 & 0 & 0\cr
-{i\over\sqrt 6}\sinh\theta_2 & -{2i\over 3}\cosh\theta_2 & 0 & 0\cr
0 & 0 & {i\over 3}\cosh\theta_2 & 0\cr
0 & 0 & 0 & 0\cr}\pmatrix{\epsilon^{(1)}\cr \epsilon^{(2)}\cr 
\epsilon^{(3)}\cr \epsilon^{(4)}\cr}}
\eqn\conifkillfive{
{\del\over \del\psi}\pmatrix{\epsilon^{(1)}\cr \epsilon^{(2)}\cr 
\epsilon^{(3)}\cr \epsilon^{(4)}\cr} =
-{i\over 6}\pmatrix{1 & 0 & 0 & 0\cr
0 & 1 & 0 & 0\cr
0 & 0 & 1 & 0\cr
0 & 0 & 0 & -3\cr}\pmatrix{\epsilon^{(1)}\cr \epsilon^{(2)}\cr 
\epsilon^{(3)}\cr \epsilon^{(4)}\cr}}
A remarkable simplification takes place on observing that a spinor
with $\epsilon^{(1)} = \epsilon^{(2)} = \epsilon^{(3)} = 0$
automatically satisfies the first four equations. Inserting this form
in the last equation, we find
\eqn\coniffinaleq{
{\del\over \del\psi}\epsilon^{(4)} = {i\over 2}\epsilon^{(4)}}
from which the corresponding Killing spinor is
\eqn\coniffinalsoln{
\epsilon = e^{{i\over 2}\psi}\pmatrix{0\cr 0\cr 0\cr 1}}
It is easy to see that there are no other solutions to the coupled set
of equations. For each of the components $\epsilon^{(1)},
\epsilon^{(2)}, \epsilon^{(3)}$, there is an irrational factor
$\sqrt{6}$ in one or other equation, which implies that we can never
get a solution that is single-valued in the angles $\phi_1,\phi_2$.

Thus, as expected, there is a single Killing spinor for this manifold,
proving explicitly that it has ${1\over 4}$-supersymmetry. The Killing
spinor has the very simple form given in Eq.\coniffinalsoln, and is
quite similar to a solution obtained in Ref.\refs\popewar\ in the
context of 7-dimensional Einstein spaces.

Because this Killing spinor depends only on $\psi$, any orbifold of
$W_{4,2}$ by an action under which spinors transform trivially (such
as a translation in the angles) will preserve it. One example is
provided by the $Z_k\times Z_{k'}$ transformation:
\eqn\rpinv{
\theta_1 \to \theta_1 + {2\pi\over k}, \quad 
\theta_2 \to \theta_2 + {2\pi\over k'}}
This introduces conical singularities into the Euclidean $AdS_2$
factors that make up the base of $W_{4,2}$, but as usual one expects
that string propagation on this space is smooth. 

\newsec{Discussion}

We have constructed Kiling spinors in various coordinate systems and
thereby discovered a number of supersymmetric orbifolds of
$AdS_5$. One should attempt to understand the global properties and
causal structure of such spacetimes. Some of them are already known to
be topological black holes.

One interesting proposal emerges from our discussion. There is an
$AdS_5/Z_2$ orbifold with a circle of fixed points, very similar to
the $S^5/Z_2$ orbifold obtained by placing D3-branes at a $Z_2$ ALE
singularity. Both are cases of $\half$-supersymmetry. Now for the
latter, it is known\refs\klebwit\ that blowing up the circle of fixed
points is a relevant deformation which causes the theory to flow to
the ${1\over 4}$-supersymmetric theory obtained by replacing $S^5/Z_2$
with the Stiefel manifold $V_{4,2}$. The corresponding conformal
theories flow from $N=2$ to $N=1$ and can be obtained very simply by
rotating branes in a brane construction\refs{\uranga,\dasm}. One could
perhaps expect an analogous blowup of our $AdS_5/Z_2$ orbifold to lead
to the non-compact Stiefel manifold $W_{4,2}$ discussed in
Ref.\refs\bsv. The physics of this would be quite different from the
compact case and possibly very interesting\foot{This proposal arose in
discussions with Debashis Ghoshal.}.

In comparing our results with those of Ref.\blust, we find that we
have reproduced the $Z$-orbifold discussed there and explicitly shown
that it is $\half$-supersymmetric, which is important for the
identification proposed with $pp$-waves on a brane. However, it is not
clear if we have found the $Z\times Z$ orbifold that also finds brief
mention in their work. This is supposed to be dual to a 3-brane with a
pp-wave together with a D-instanton, and one would expect it to be
${1\over 4}$-supersymmetric. We have found two ${1\over
4}$-supersymmetric orbifolds that include the action in
Eq.\lightorb. One is found by adjoining the $Z$ action which
compactifies one or more of the spatial coordinates in the brane, as
discussed below Eq.\secondcase. The other is obtained by adjoining a
$Z_k$ action, as discussed after Eq.\commspin. In neither of these
cases does the second orbifold group appear symmetrically with the
first, while the authors of Ref.\refs\blust\ seem to suggest that such
an example should exist. We hope to return to this point in the
future, along with a study of the physical interpretation, in terms of
the brane worldvolume theory, for the various orbifolds we have
constructed here.

The compactification of type IIB string theory on $AdS_5\times S^5$
possesses a remarkable ``symmetry'' between the two
factors. Geometrically, the only difference is that while $S^5$ solves
a quadratic equation in $R^6$, $AdS_5$ solves a quadratic equation in
$R^{4,2}$. This ``symmetry'' is not reflected in the emergence of this
background as the near-horizon geometry of D3-branes, where $R^6$ has
a physical interpretation as the flat space transverse to the
3-branes, but $R^{4,2}$ does not appear. One may speculate that such a
symmetry may be visible or partly visible in F-theory, which can
sometimes be given a 12-dimensional interpretation. This philosophy
has been partially explored in Ref.\refs\blust\ using earlier
observations in Ref.\refs\tseytlin. It would be interesting to find
out whether such a symmetry can be exploited to systematically
classify the supersymmetric orbifolds of $AdS_5$.
\bigskip

\noindent{\underbar{\bf Acknowledgements:} }

We are grateful to Sumit Das, Justin David, Debashis Ghoshal and
Nemani V. Suryanarayana for helpful comments.

\listrefs

\end